\documentclass{PoS}
\def\ltap{\ \raise.3ex\hbox{$<$\kern-.75em\lower1ex\hbox{$\sim$}}\ }

\usepackage{bm}

\title{Theoretical analysis of $\Lambda$(1405) photoproduction}

\ShortTitle{$\Lambda$(1405) photoproduction}

\author{\speaker{Satoshi X. Nakamura}\\
        Department of Physics, Osaka University, Toyonaka, Osaka 560-0043, Japan\\
        E-mail: \email{nakamura@kern.phys.sci.osaka-u.ac.jp}}

\author{{Daisuke Jido}\\
        Department of Physics,
Tokyo Metropolitan University,
Hachioji, Tokyo 192-0397, Japan\\
	}

\abstract{
We develop a model that describes 
the $\gamma p \to K^{+} \pi \Sigma$ reaction 
in the $\Lambda(1405)$ region.
The model consists of gauge invariant photo-production mechanisms, and
the chiral unitary model that gives
the rescattering amplitudes where $\Lambda(1405)$ is contained.
The model also contains 
phenomenological parameters, associated with short-range dynamics,
to be used in fitting data.
We successfully fit
recent CLAS data for the $\pi\Sigma$ invariant mass distributions
(line-shape)
in the $\gamma p \to K^{+} \pi \Sigma$ reaction
for all the charge states.
We find that the higher mass pole for $\Lambda(1405)$ of the chiral
unitary model plays an important role in the reaction.
We also find the nonresonant background
contribution is not negligible, and its sizable effect shifts
the $\Lambda(1405)$ peak position by several MeV.
This work sets a starting point for a fuller analysis in which 
line-shape as well as $K^+$ angular distribution data
are simultaneously analyzed for
extracting $\Lambda(1405)$ pole(s).
}

\FullConference{XV International Conference on Hadron Spectroscopy-Hadron 2013\\
		4-8 November 2013\\
		Nara, Japan }

\begin{document}

\section{Introduction}
\label{sec:intro}

Recently, the CLAS collaboration at Jefferson Laboratory conducted
a high statistics, wide angle coverage experiment for
the $\gamma p \to K^{+} \pi \Sigma$ reaction
for center-of-mass energies $1.95 < W < 2.85$ GeV~\cite{Moriya:2013eb,Moriya:2013hwg}.
In this experiment, all the three charge states of the $\pi \Sigma$
channels were simultaneously observed in the $\gamma p$ scattering
for the first time, 
and the differential cross sections were measured 
for the $\pi\Sigma$ invariant mass distribution (line-shape)
and for the $K^+$ angular distribution.
This is the cleanest data that cover the kinematics of $\Lambda(1405)$
excitation, and it is interesting to examine if 
the $\Lambda(1405)$ pole(s) can be extracted from the data for the first time.

The pole structure of the $\Lambda(1405)$ resonance is a key issue to understand 
the nature of $\Lambda(1405)$ and the $\bar KN$ interaction. 
The coupled-channel approach based on 
the chiral effective theory (chiral unitary model) suggests that the $\Lambda(1405)$ resonance
is composed by two poles located between the $\bar KN$ and $\pi \Sigma$ 
thresholds~\cite{Hyodo:2011ur} and these states have different masses, widths
and couplings to the $\bar KN$ and $\pi\Sigma$ channels. One pole 
is located at $1426- 16i$ MeV with a dominant coupling to $\bar KN$, while
the other is sitting at $1390-66i$ MeV with a strong coupling to $\pi \Sigma$~\cite{Jido:2003cb}.
These two states are generated dynamically by the attractive interaction in 
the $\bar KN$ and $\pi\Sigma$ channels with $I=0$ ($I$: total isospin)~\cite{Hyodo:2007jq}.
Because the $\Lambda(1405)$ resonance is composed by two states 
which have different weight to couple with $\bar KN$ and $\pi \Sigma$,
the spectral shape of the $\pi\Sigma$ line-shape in the $\Lambda(1405)$ region
depends on how $\Lambda(1405)$ is produced, as pointed out in Ref.~\cite{Jido:2003cb}.  

It is important to confirm the two-pole structure by analyzing the new
CLAS data for $\gamma p \to K^{+} \pi \Sigma$, and if so,
it will be interesting to see how the two-pole structure plays a role in
the $\pi\Sigma$ line-shape.
In order to extract the $\Lambda(1405)$ resonance pole(s) from
the production data, one needs to develop a model that consists of production
mechanism followed by the final state interaction (FSI);
$\Lambda(1405)$ is excited in the FSI.
Through a careful analysis of the data, one can pin down 
the production mechanism as well as the scattering amplitude
responsible for the FSI.
Then the $\Lambda(1405)$ pole(s) will be extracted from the scattering amplitude.
In this work, we consider production mechanisms that are gauge invariant
at the tree-level.
We consider relevant meson-exchange mechanisms, and contact terms that
simulate short-range mechanisms. 
For the rescattering amplitude that contains $\Lambda(1405)$, we use the
chiral unitary model.
We successfully fit the CLAS data with it. 
Then we discuss a role played by each mechanism, effects of non-resonant
contributions.
By doing so, we set a starting point for a full analysis in which we
simultaneously analyze the data for line-shape~\cite{Moriya:2013eb}
and the $K^+$ angular distribution~\cite{Moriya:2013hwg} to study the
pole structure of $\Lambda(1405)$.
Details of this work, including more elaborate description of the model and more
results, are reported in our recent publication~\cite{NJ}.

\section{Model}
\label{sec:model}

We describe the $\gamma p \to K^+\pi\Sigma$ reaction by a set of
tree-level mechanisms for $\gamma p \to K^+M_jB_j$ 
($M_jB_j$ : a set of meson and baryon)
followed by $M_jB_j\to \pi\Sigma$ rescattering, where
$M_jB_j=K^-p,\bar K^0 n$, $\pi^0\Lambda, \pi^0\Sigma^0, \eta\Lambda, \eta\Sigma^0$,
$\pi^+\Sigma^-, \pi^-\Sigma^+, K^+\Xi^-, K^0\Xi^0$,  respectively.
Thus the reaction amplitude for the $\gamma p \to K^+M_jB_j$ reaction
is given by 
$T^j = V^j + T^j_{R}$, 
where $V^j$ is the set of tree-level photo-production mechanisms that we discuss
in the next paragraph.
Contribution from the rescattering is denoted by $T^j_{R}$.
The rescattering amplitude is calculated with a partial wave expansion
with respect to the relative motion of $M_jB_j$, and 
$(J,L)=(1/2,0)$ and $(1/2,1)$ partial waves are considered; $J$ and $L$
are the total and orbital angular momenta for $M_jB_j$.
The partial wave amplitude is given, with the on-shell factorization, by
$T^j_{R;JL} = \sum_{j'}\, T^{jj'}_{JL}\, G^{j'}\, V^{j'}_{JL}$
where 
$T^j_{R;JL}$ and $V^{j}_{JL}$
are partial wave amplitudes of
$T^j_{R}$ and $V^j$, respectively, 
and are calculated with the on-shell momenta of relevant particles.
For the $M_{j'}B_{j'}\to M_jB_j$ scattering amplitudes $T^{jj'}_{JL}$,
we use those from the chiral unitary model
given in Ref.~\cite{ORB} for $(J,L)=(1/2,0)$ wave, and 
in Ref.~\cite{JOR} for $(J,L)=(1/2,1)$ wave.
The $(J,L)=(1/2,0)$ wave contains $\Lambda(1405)$ as double poles, 
while the $(J,L)=(1/2,1)$ wave
provides a smooth background.
We use the meson-baryon Green function, $G^j$, calculated with the
dimensional regularization.
The subtraction constants contained in $G^j$
can depend on a channel $j$ as well as a
production mechanism contained in $V^j$.

We consider gauge-invariant tree-level photo-production mechanisms 
($V^j$) as follows:
minimal substitution to the lowest order chiral meson-baryon interaction
such as the Weinberg-Tomozawa terms and the Born
terms;
vector-meson exchange mechanisms.
These photo-production mechanisms are expanded in terms of $1/M_B$,
and ${\cal O} (1)$ and ${\cal O} (1/M_B)$ terms are taken in our calculation.

With the meson-exchange production mechanisms
and the subtraction constants
taken as the same as those in the chiral unitary amplitudes, 
we cannot reproduce the $\pi\Sigma$ line-shape data
for the $\gamma p\to K^+\pi\Sigma$ reaction
from the CLAS~\cite{Moriya:2013eb}.
Therefore, it is inevitable to introduce adjustable degrees of freedom
to fit the data.
Thus all of the meson-exchange mechanisms $V^j$ are multiplied by a
common dipole form factor, and 
the cutoff is fitted to the data.
In addition, we also consider
phenomenological contact terms that can simulate mechanisms not
explicitly considered, such as, in particular, $N^*$ and $Y^*$ excitation mechanisms.
We take couplings for the contact terms $W$-dependent ($W$ : total energy of the system),
and will be determined by fitting the $\gamma p\to K^+\pi\Sigma$ data~\cite{Moriya:2013eb}.
The subtraction constants
are also adjusted to fit the data, thereby changing the interference
pattern between different production mechanisms.
It is noted that we do not adjust the subtraction constants in the
chiral unitary amplitudes in the fit.
The subtraction constants we adjusted are all for the first loop of the
rescattering, and for the renormalization of the production mechanism.
By introducing quite a few fitting parameters, 
our method could bring a model-dependence 
when we extract $\Lambda(1405)$ pole(s) from the data.
The model-dependence of $\Lambda(1405)$ pole(s) 
must be assessed by analyzing the data with
different form factors and/or contact terms.
This will be a future work.

\begin{figure}
\includegraphics[clip,width=0.45\textwidth]{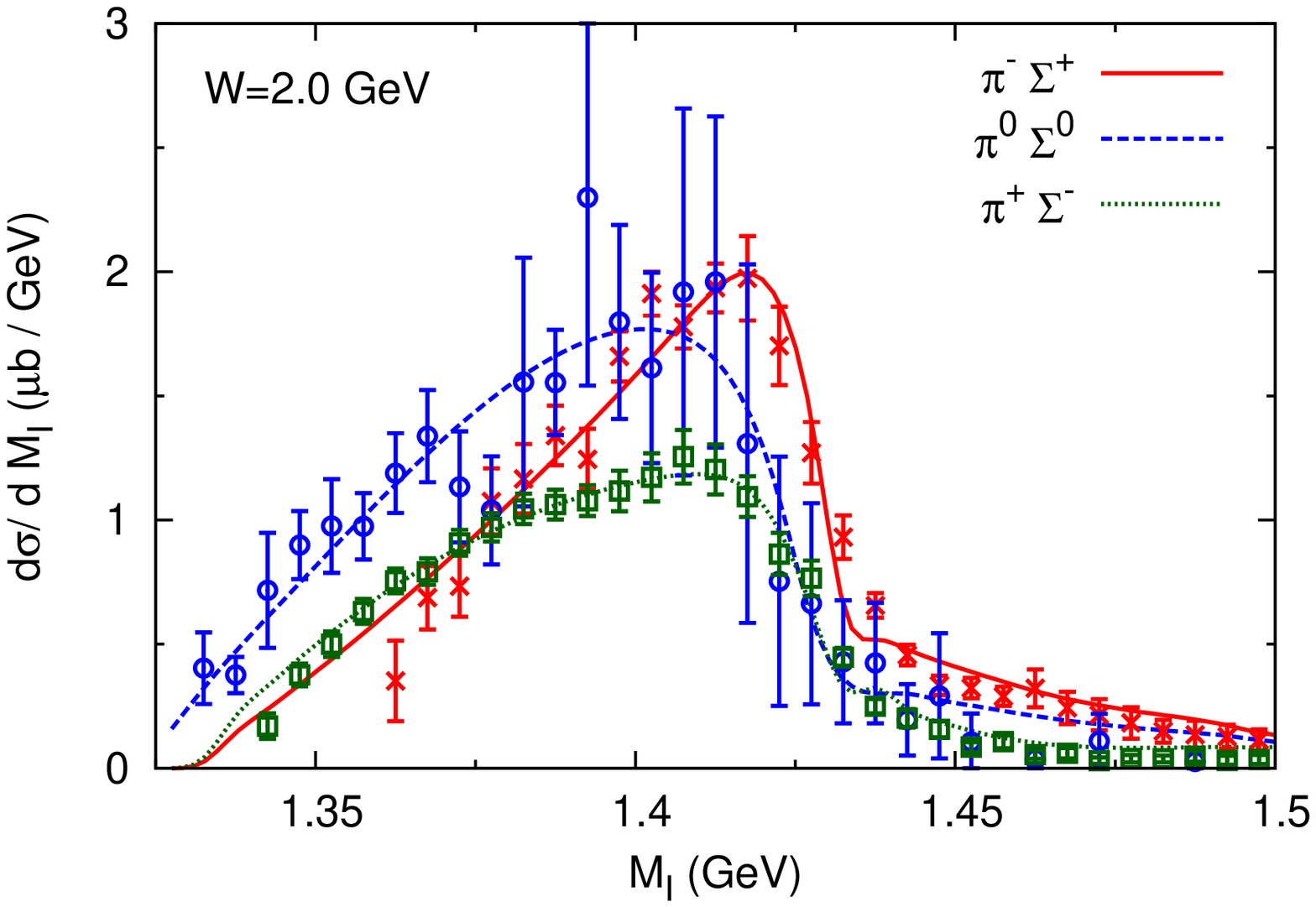}
\includegraphics[clip,width=0.45\textwidth]{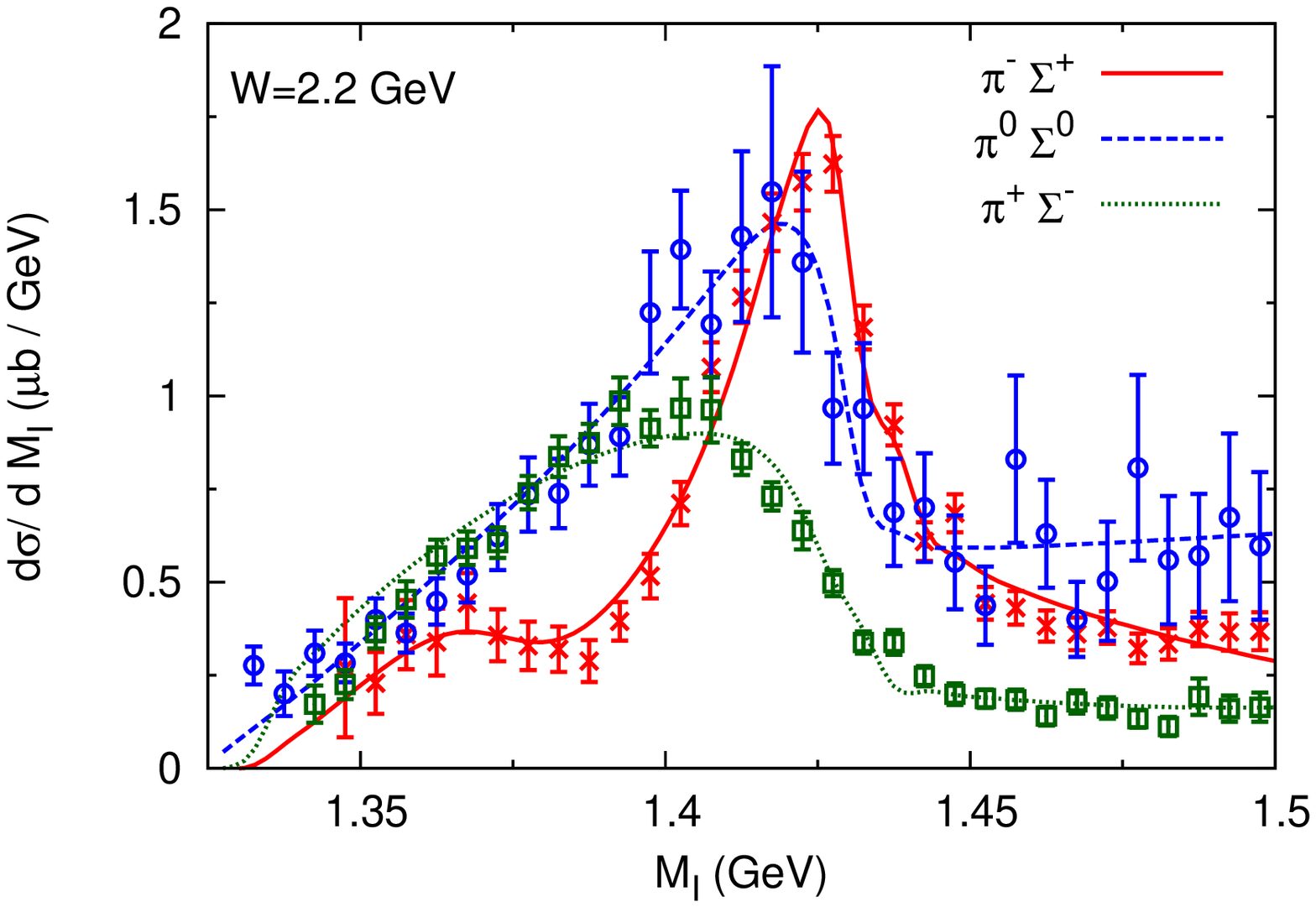}
\caption{\label{fig:2gev} (Color online)
Comparison of $\pi\Sigma$ line-shapes from our model with
 data~\cite{Moriya:2013eb} at $W=2.0$, 2.2~GeV.
Symbols for the data are cross (red) for $\pi^-\Sigma^+$,
circle (blue) for $\pi^0\Sigma^0$, and
square (green) for $\pi^+\Sigma^-$.
}
\end{figure}

\section{Result}
\label{sec:result}

We show the $\pi\Sigma$ line-shapes from our model
in Fig.~\ref{fig:2gev} where
the CLAS data are also shown for comparison. 
Our model fits the data very well for all three
different charge states of $\pi\Sigma$.

\begin{figure}
\includegraphics[clip,width=0.45\textwidth]{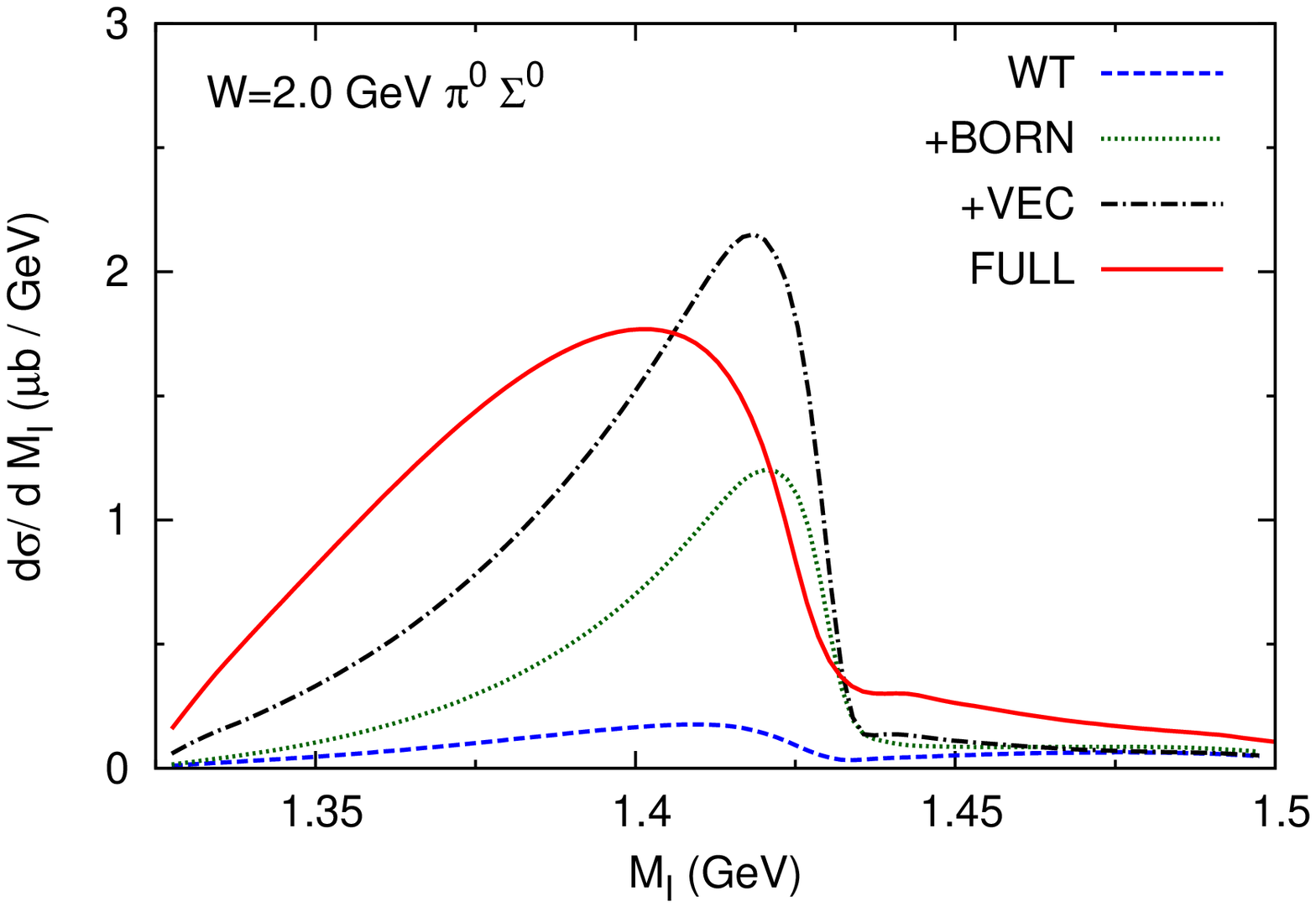}
\includegraphics[clip,width=0.45\textwidth]{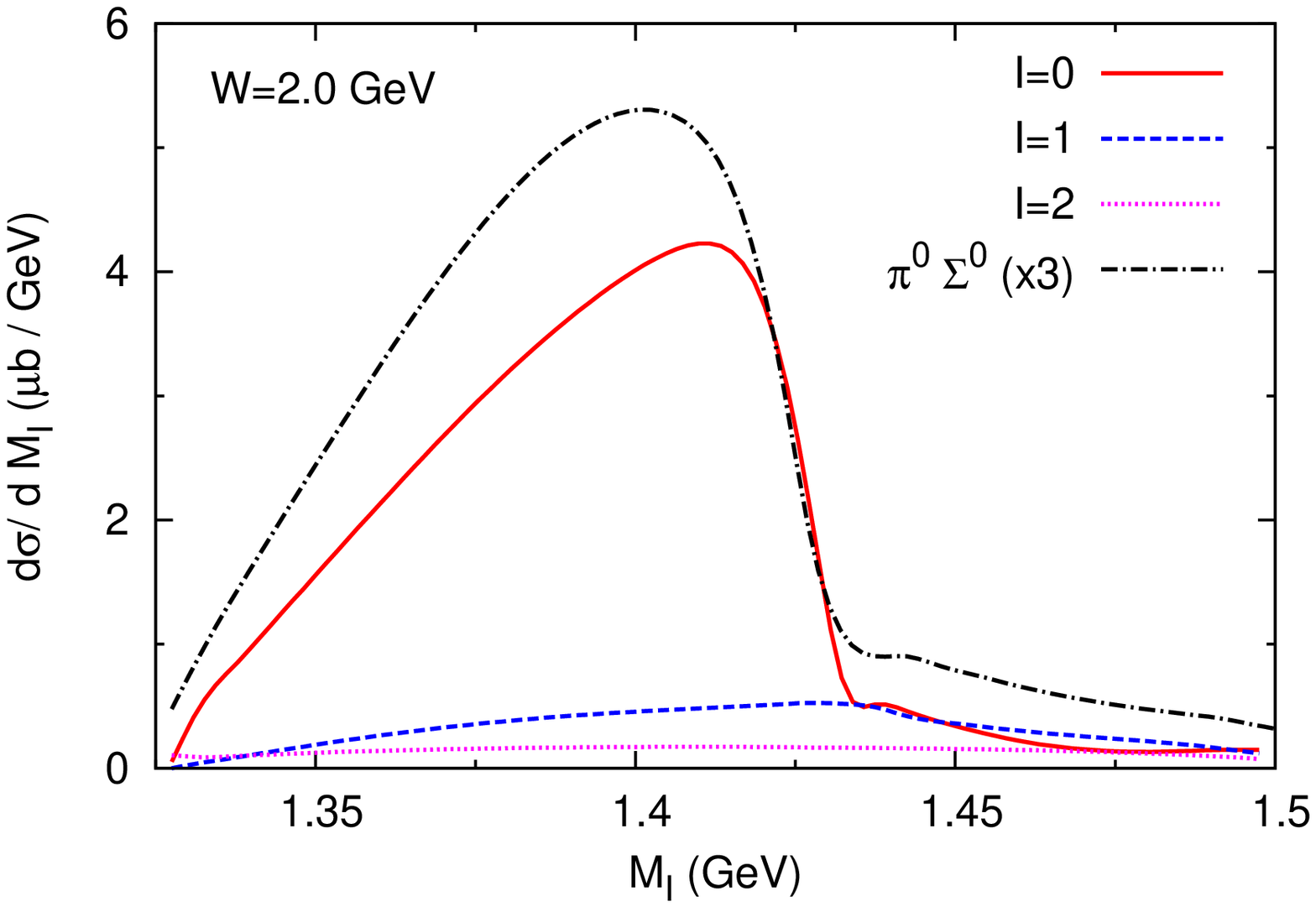}
\caption{\label{fig:mec} (Color online) (LEFT)
Contribution of each production mechanism for 
$\gamma p\to K^+\pi^0\Sigma^0$.
Contribution from the gauged Weinberg-Tomozawa terms (WT)
is given by the blue dashed line.
Contribution that additionally includes the gauged Born terms (+BORN)
is given by the green dotted line.
Contribution that further includes the vector-meson exchange
 (+VEC) is given by the black dash-dotted line.
The full result with the contact terms
is shown by the red solid line.
(RIGHT) Isospin decomposition of $\pi\Sigma$ line-shapes. Contributions from the
 isospin states $I$ are shown, along with the $\pi^0\Sigma^0$ line-shape
  multiplied by 3.
}
\end{figure}

It is interesting to break down the line-shapes into 
contributions from different mechanisms, as shown in Fig.~\ref{fig:mec} (left).
As seen in the figure, different mechanisms give significant
contributions that interfere with each other.
We find that the contributions from the gauged Weinberg-Tomozawa terms
are rather small,
as a result of a destructive interference between several 
gauged Weinberg-Tomozawa terms.
This destructive interference is not necessarily a result of the gauge
invariance, but rather
relevant subtraction constants have been fixed by the fit
so that the cancellation happens.
Meanwhile, the contact terms, which simulate short-range dynamics,
also give a large contribution to bring the theoretical calculation
into agreement with the data. 
Finally, we mention that coupled-channels effects are mostly from 
the $\bar K N$ and $\pi\Sigma$ channels.
The difference in the line-shape
between different charge states observed in Fig.~\ref{fig:2gev}
is a result of the interference between different isospin states.
The $\pi\Sigma$ has three isospin states ($I=0,1,2$), and they are
separately shown in Fig.~\ref{fig:mec} (right).
A dominant contribution is from the $I=0$ state as expected due to the
$\Lambda(1405)$ peak. 
The higher mass pole at $1426- 16i$ MeV, that creates the prominent bump
in the line-shapes, seems to play more important role than the lower
mass pole. 
This is because the production mechanisms in our model generate $\bar K N$ more strongly
than $\pi\Sigma$, and the final state interaction induces 
$\bar K N\to \pi\Sigma$.
As shown in the previous study~\cite{Jido:2003cb}, the higher mass pole 
couples to the $\bar K N$ channel more strongly than the lower mass pole
does. 
The $I=1$ state gives a smaller contribution, but still plays an important
role to generate the charge dependence.
The $I=2$ state contribution is even smaller, but still
non-negligible.
To see this point, we show in Fig.~\ref{fig:mec} (right)
the $\pi^0\Sigma^0$ line-shape multiplied by 3.
The difference between this and the $I=0$ line-shape is the effect of
the interference between the $I=0$ and $I=2$ states. 
We can see that the interference with the $I=2$ state even changes slightly the peak
position of the $\pi^0\Sigma^0$ line-shape.

\begin{figure}
\includegraphics[clip,width=0.45\textwidth]{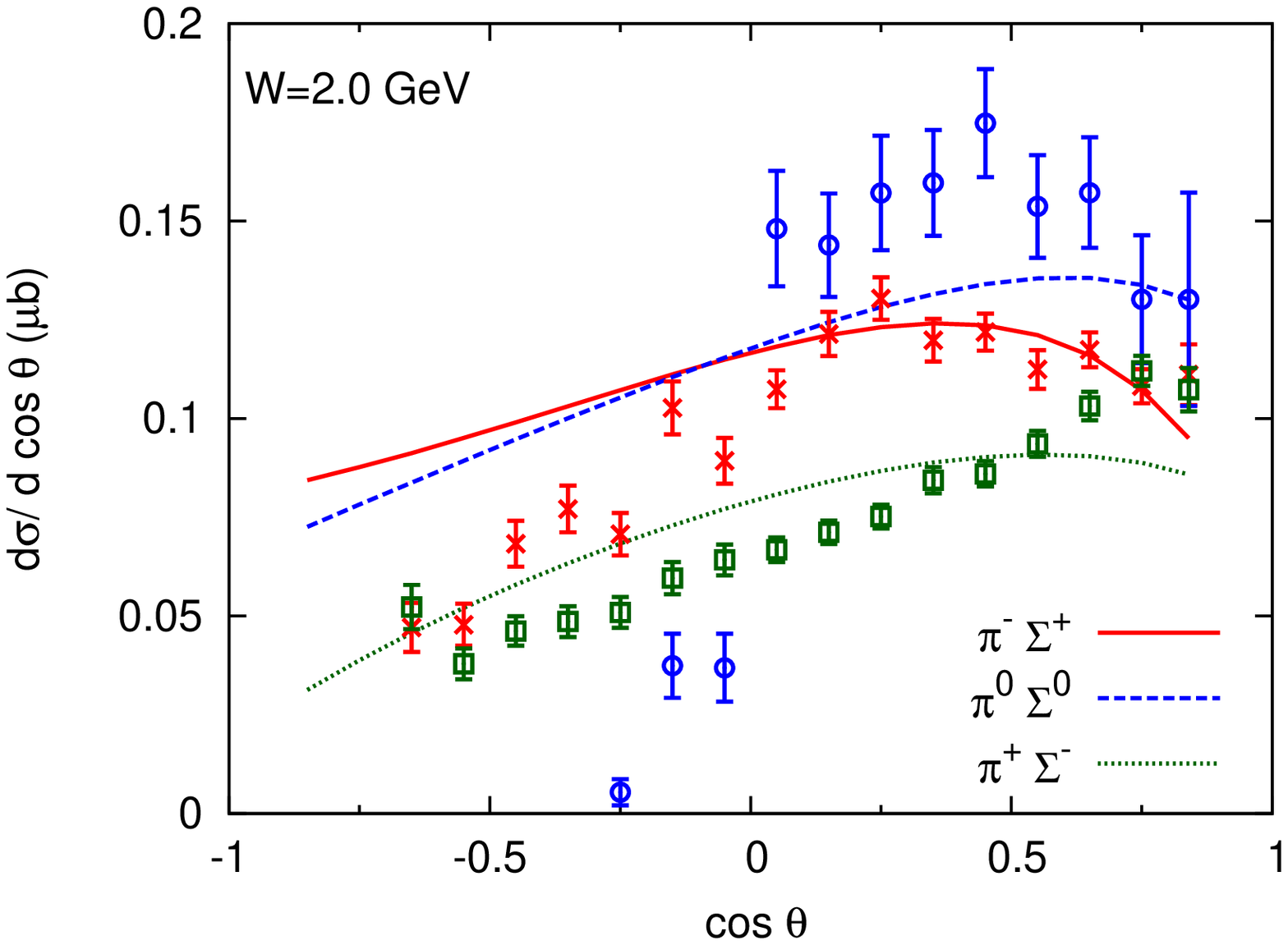}
\includegraphics[clip,width=0.45\textwidth]{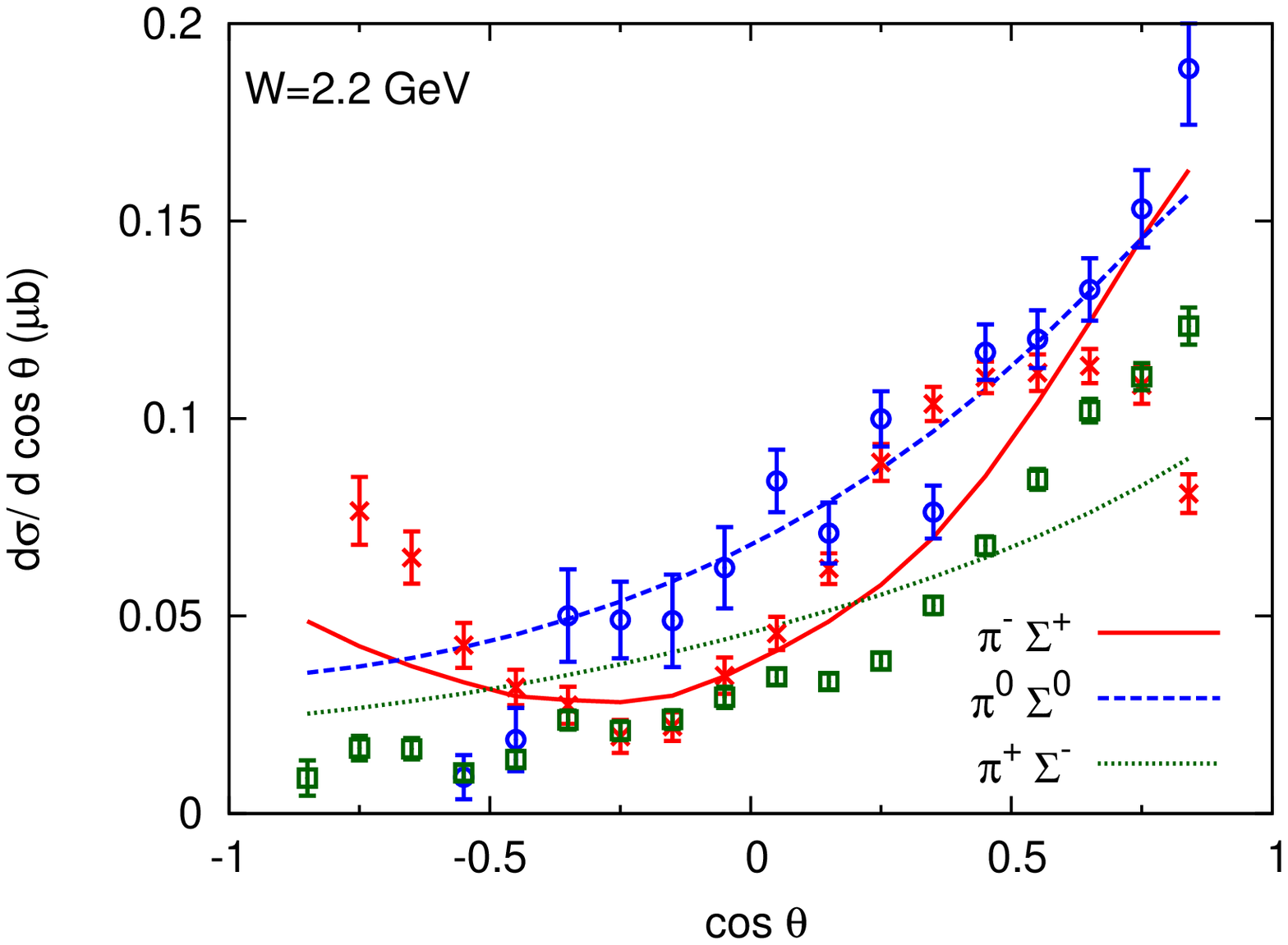}
\caption{
\label{fig:angle} (Color online)
Comparison of the $K^+$ angular distributions for $\gamma p\to K^+\pi\Sigma$ 
at $W=2.0, 2.2$~GeV with data from the CLAS~\cite{Moriya:2013hwg}. 
}
\end{figure}

Fitting only the $\pi\Sigma$ line-shape data, 
we found 
several solutions whose quality of the fit to the line-shape data are
comparable. However, they can have very different $K^+$ angular distribution.
Therefore, $K^+$ angular distribution data will be
useful information to constrain the production mechanism.
Recently the CLAS Collaboration reported data for the $K^+$ angular
distributions~\cite{Moriya:2013hwg}. 
Here we show in Fig.~\ref{fig:angle} the $K^+$ angular distributions from our model that
reproduces the data relatively better than the other solutions.
At $W=2.2$~GeV, our model captures overall trend of the data. 
However, for the $\gamma p\to K^+\pi^0\Sigma^0$ reaction at $W=2.0$~GeV, 
there is a sharp rise in the data at $\cos\theta\sim 0$ while rather
smooth behavior is found in the calculated counterpart. 
We actually tried fitting the $K^+$ angular distributions data, but this
sharply rising behavior cannot be fitted with the current setup. 
It seems that we need to search for a mechanism that is responsible for
this behavior. 
We leave such a more detailed analysis of the $K^+$ angular distribution
to a future work.

\begin{acknowledgments}
SXN is the Yukawa Fellow and his work is supported in part by Yukawa
Memorial Foundation.
This work was partially supported by the Grants-in-Aid for Scientific Research 
(No.\ 24105706).
\end{acknowledgments}

\end{document}